\documentclass[amsmath,amssymb,amsfonts,twocolumn]{revtex4}
\usepackage{graphicx}

\def \beq {\begin{equation}}
\def \eeq {\end{equation}}
\def \tr {\rm Tr}

\begin{document}
\title{Magnetic Sensitivity and Entanglement Dynamics of the Chemical Compass}
\author{I. K. Kominis}

\affiliation{Department of Physics, University of Crete, Heraklion 71103, Greece}

\begin{abstract}
We present the quantum limits to the magnetic sensitivity of a new kind of magnetometer based on biochemical reactions.
Radical-ion-pair reactions, the biochemical system underlying the chemical compass, are shown to offer a new and unique physical realization of a magnetic field sensor competitive to modern atomic or condensed matter magnetometers. We elaborate on the quantum coherence and entanglement dynamics of this sensor, showing that they provide the physical basis for testing our understanding of the fundamental quantum dynamics of radical-ion-pair reactions.
\end{abstract}
\maketitle
Quantum physics, in particular quantum information, control and measurement are concepts rapidly infiltrating biological or biochemical processes. For example, quantum coherence has been shown to play a fundamental role in photosynthesis \cite{engel,lloyd,johnsonNF,plenio,pan,caruso,collini,scholes}, while quantum measurement dynamics have been demonstrated to underlie radical-ion-pair reactions \cite{kom1,JH,kom2}, the spin-dependent biochemical reactions at the heart of the avian magnetic compass mechanism. Even electron spin entanglement has also been addressed with respect to the chemical compass \cite{briegel,vedral}, paving the way for the dawn of quantum biology \cite{ball}.

We will here show that spin-selective radical-ion-pair reactions are no different than atomic \cite{budker} or solid state quantum sensors \cite{lukin} used in e.g. precision metrology \cite{huelga,geremia,andre}, and in principle are able to offer an exquisite magnetic sensitivity. We will establish the fundamental quantum limits to the magnetic sensitivity of these biochemical sensors, and we will elaborate on the fundamental decoherence mechanism present. We will show that the mechanism damping singlet-triplet coherence also damps any electron spin entanglement possibly present, as is well understood in precision measurements. In so doing, we will also show that the recently appeared entanglement considerations \cite{briegel,vedral} are questionable, since these works have not taken into account the fundamental singlet-triplet decoherence process. Finally, we will compare the three different and currently competing master equations attempting to describe the quantum dynamics of radical-ion-pair reactions. Based on their ability to correctly account for the coherence and entanglement dynamics of these reactions, it will be shown that one out of the three theories can be eliminated. 

Radical-ion pairs (Fig. 1) \cite{ks,steiner,rodgers} are biomolecular ions with two unpaired electrons and any number of magnetic nuclei, created by a charge transfer from a photo-excited D$^*$A donor-acceptor molecular dyad DA. The magnetic nuclei of the donor and acceptor molecules couple to the two electrons via the hyperfine interaction, leading to singlet-triplet (S-T) mixing, i.e. a coherent oscillation of the total electron spin state, also affected by the electrons' Zeeman interaction with the external magnetic field. Singlet-triplet coherence (and its relaxation) is studied in a variety of contexts, as e.g. in NMR \cite{bargon,levitt,levitt2} and quantum dots \cite{johnson,marcus,koppens}. The additional complication of radical-ion-pair reaction dynamics is the spin-dependent loss of radical-ion pair population due to the recombination effect. Charge recombination terminates the reaction leading to the neutral reaction products. Angular momentum conservation at this step enforces the reaction's spin selectivity: only singlet state radical-ion pairs can recombine to reform the neutral DA molecules, whereas triplet radical-ion pairs recombine to a different metastable triplet neutral product. If the radical-ion pair's lifetime is large enough, the minute Zeeman interactions can significantly affect the reaction yields, rendering the reaction a biochemical magnetometer.
\begin{figure}[h]
\includegraphics[width=8.5 cm]{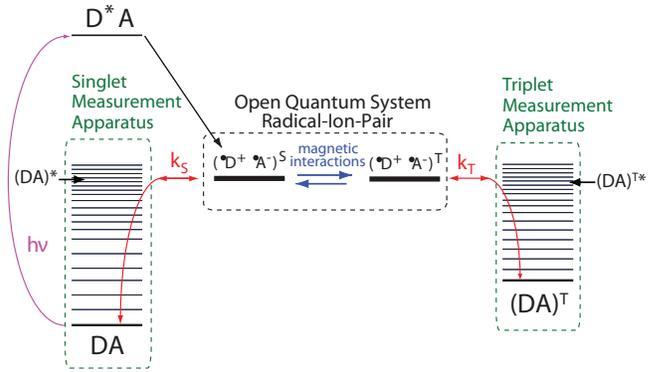}
\caption{(Color Online) A donor-acceptor dyad is photoexcited to the state D$^{*}$A, which after charge separation produces the singlet radical-ion pair $({\rm D}^{\bullet +}{\rm A}^{\bullet -})^{\rm S}$. Magnetic interactions within the molecule induce an S-T interconversion, terminated by the recombination event, that results into a neutral singlet, DA or triplet product, (DA)$^{\rm T}$. The fundamental decoherence process is the S-T decoherence, due to the coupling to the excited vibrational levels of DA and (DA)$^{\rm T}$, denoted by (DA)$^*$ and (DA)$^{T*}$, respectively. These perform two functions: (i) virtual transitions from the radical-ion-pair to those levels and back interrupt the unitary magnetic evolution, causing S-T decoherence at a rate $(k_{S}+k_{T})/2$, and (ii) they are a sink of radical-ion-pair population, as real transitions to them result in the radical-ion pair's singlet (triplet) recombination at a rate $k_{S}$ ($k_{T}$).}
 \label{fig1}
\end{figure}
The actual physical signal carrying the magnetic field information can take various forms, for example it can be the reaction yield e.g. the singlet, it can be the magnetic-field-dependent time evolution of (i) the radical-ion pair population, measured e.g in optical absorption experiments \cite{maeda}, or (ii) the singlet radical-ion-pair population measured e.g. in fluorescence experiments \cite{mfe}. No matter what the actual magnetometric signal or the particular measurement scheme, the fundamental magnetic sensitivity of this sensor is determined by general arguments no different than similar considerations in atomic sensors, i.e. based on the energy resolution during a finite measurement time, in this case the reaction time. Indeed, it is well known \cite{budker,huelga,geremia,andre} that if we use a {\it single} quantum system to measure an energy splitting $E$ during time $T$, Heisenberg's uncertainty relation (with $\hbar=1$) limits the precision $\delta E$ to $\delta E=1/T$. It is also known that we can do better than that by simultaneously employing $N$ (uncorrelated) quantum systems, leading to an improvement by $\sqrt{N}$. No matter how the measurement is performed, the fundamental shot-noise-limited energy measurement precision is thus $\delta E=1/\sqrt{N}T$. Now, if we let an ensemble of electron spins interact with a magnetic field $B$ for a time $T$, the limit $\delta E$ readily translates into the limit $\delta B$, the smallest measurable change of the magnetic field, or equivalently, the magnetic sensitivity, which is $\delta B=\delta E/\gamma$, where $\gamma$=2.8 MHz/G is the electron gyromagnetic ratio. So the fundamental magnetic sensitivity of a system employing $N$ electron spins is $\delta B=1/\gamma\sqrt{N}T$.

Radical-ion pair (RP)  reactions start by photo exciting the neutral precursors. We assume that at time $t=0$ a flash of light creates  $N_{0}$ radical pairs. The RP reaction, i.e. the creation of the neutral reaction products, takes place at a characteristic time $T_{r}$, the reaction time. Considering for simplicity a mono-exponential decay of the radical-ion-pair population of the form $N(t)=N_{0}e^{-t/T_{r}}$, the 
maximum of $t\sqrt{N(t)}$ (and hence the minimum of $\delta B$) is realized at $t=2T_{r}$, hence, apart from order-of-unity factors, the shot-noise-limited magnetic sensitivity is $\delta B={1/{\gamma\sqrt{N_{0}}T_{r}}}$.
If the photoexcitation-reaction cycle is repeated $n$ times for a total measurement time $\tau=n(2T_{r})$, averaging the $n$ measurements will yield 
a $\sqrt{n}$ improvement, thus
\beq
\delta B={1\over \gamma}{1\over \sqrt{N_{0}\tau T_{r}}}.\label{deltaB}
\eeq
For an order of magnitude  estimate it is noted that concentrations of radical-ion pairs on the order of  $10^{16}~{\rm cm}^{-3}$ can be obtained, while reaction times up to 10 $\mu$s or higher are not uncommon, leading, for an active volume of 1 cm$^{3}$, to a sensitivity on the order of $\delta B\approx 10^{-13}/\sqrt{\tau[s]}~{\rm G}/\sqrt{\rm Hz}$. The bandwidth of the magnetometer will obviously be on the order of $1/\tau$.

All quantum sensors are plagued by decoherence, this one being no exception. As explained in detail in \cite{kom1,kom2}, the fundamental decoherence process of this magnetometer stems from the continuous measurement of the electron spin state performed by the coupling of the radical-ion-pair to the reservoir states, which are inherent in the molecule DA, as shown in Fig. 1. This coupling results in projections on the singlet or triplet subspace, occurring randomly along the singlet-triplet (S-T) mixing, and leading to the damping of S-T coherence. This is an intrinsic, minimum and unavoidable, or in other words fundamental decoherence process, any other extrinsic spin relaxation phenomena only adding up and reducing the sensitivity (making $\delta B$ larger). The relevant decoherence time turns out to be \cite{kom1,kom2} on the order of the reaction time $T_r$. This is why the decoherence time does not independently enter \eqref{deltaB}, in contrast to  atomic magnetometers \cite{budker}, the sensitivity of which is determined by two unrelated time scales, the measurement time and the decoherence time. Here these two time scales coincide. It is also noted that in current biological/biochemical realizations the actual sensitivity is far from saturating the limit \eqref{deltaB}, the main reason being the sub-optimal magnetometric signal measurement, which is far from being shot-noise-limited \cite{weaver,maeda}. In that case, i.e. if the measurement's signal-to-noise ratio is ${\cal S/N}\leq\sqrt{N_{0}}$, and setting henceforth $\tau=T_{r}$, it follows that
\beq
\delta B={1\over {\gamma({\cal S/N})}}{1\over T_{r}}.\label{msens}
\eeq
The preceding discussion is well-understood in the field of precision measurements, for which these quantum limits are what the impossibility of perpetuum mobile constructs is in the field of thermodynamics. For the skeptic general reader, however, we will give a concrete example elucidating the fundamental impasse posed by such limits.
One could ask, if the singlet reaction yield $Y_{S}=Y_{S}(B)$, clearly dependent on $B$, is chosen as the magnetometric signal, why is the precision $\delta B$ limited by the reaction time? The answer, as mentioned in the introduction, is that the shorter the reaction time, the less dependent the yield on the magnetic field. This can be easily seen, because in the limit $T_{r}\rightarrow 0$ it would be $Y_{S}\rightarrow 1$, i.e. the magnetic Hamiltonian has no time to mix the electron spin, hence the reaction products are all singlet (if the initial state was a singlet), independently of $B$. In the general case, and assuming equal recombination rates $k_{S}=k_{T}=k$, it can be shown \cite{timmel} that e.g. the singlet reaction yield is $Y_{S}=(1/M)\sum_{n=1}^{4M}\sum_{m=1}^{4M}|(Q_{S})_{nm}|^{2}k^{2}/(k^{2}+\omega_{nm}^{2})$, where $M$ is the nuclear spin multiplicity, $\omega_{n}$ are the eigenvalues of ${\cal H}$ with $n=1,2,...,4M$ (4 is the multiplicity of the two-electron spin space), $\omega_{nm}=\omega_{n}-\omega_{m}$, $(Q_{S})_{nm}$ is the matrix element of $Q_{S}$ in the eigenbasis of ${\cal H}$ and $Q_{S}$ the 4M-dimensional singlet projection operator. Thus, as $k\rightarrow\infty$, the sensitivity, $dY_{S}/dB$, of the yield on the magnetic field, encoded in $\omega_{nm}$, tends to zero as $1/k^{2}$. If the precision in measuring $Y_{S}$ were $\delta Y_{S}$, then the magnetic sensitivity would be $\delta B=\delta Y_{S}/(dY_{S}/dB)$ and would scale as $1/T_{r}^{2}$, over-satisfying the limit \eqref{msens}.

The limit \eqref{msens} was derived on the usual assumption of uncorrelated single-spin systems. Radical-ion pairs, however, are two-electron-spin systems, a possible entanglement of which could in principle lead to an improvement in $\delta B$ by a factor of $\sqrt{2}$ at most, as already known \cite{huelga,geremia,andre}. Neglecting this order-of-unity enhancement, we repeat that \eqref{msens} is a hard limit that cannot be surpassed, no matter how ingenious the measurement scheme. Generalizing the above example, if we choose to measure an observable, ${\cal O}={\cal O}(B)$, which depends on the magnetic field, and the measurement precision of ${\cal O}$ is $\delta{\cal O}$, then the measurement precision of $B$ will be
\beq
\delta B_{\cal O}={{\delta{\cal O}}\over {|d{\cal O}/d B |}},\label{sens}
\eeq
where $|d{\cal O}/dB|$ is the sensitivity of the B-dependence of the observable ${\cal O}$. The achieved limit $\delta B_{\cal O}$ cannot be lower than $\delta B$ no matter what, i.e. $\delta B_{\cal O}/\delta{B}\geq 1$.
Recently, however, in their study of the entanglement dynamics of radical-ion-pair reactions, Briegel and coworkers \cite{briegel} found a fundamentally different result, severely violating the above inequality.
These authors introduced an observable ${\cal O}=T_{E}$, the entanglement lifetime, shown to have a steep dependence on the magnetic field $B$, depicted in Fig. 2b of \cite{briegel}. We can use this steep dependence to measure $B$. To that end we would also need the precision $\delta T_{E}$ of measuring the entanglement lifetime $T_{E}$. Measuring time is like measuring frequency, or energy. If we measure a frequency $f$ during $T_{r}$, the best possible precision is $1/T_{r}$, again improved by $(S/N)$, i.e. $\delta f=1/T_{r}(S/N)$. If $f=1/T_{E}$, then $\delta f=\delta T_{E}/T_{E}^{2}$, hence
$\delta T_{E}=T_{E}^{2}/[T_{r}(S/N)]$. Using \eqref{sens} we find that the magnetic sensitivity attained through a measurement of $T_{E}$ is
\beq
\delta B_{T_{E}}={1\over {(S/N)}}{T_{E}^{2}\over T_{r}}{1\over {|dT_{E}/dB|}}.\label{sensB}
\eeq
It must be $\delta B_{T_{E}}\geq \delta B$, thus the inequality $|dT_{E}/dB|\leq \gamma T_{E}^{2}$ must be satisfied, otherwise the fundamental limit \eqref{msens} is violated. However, in the calculation of the entanglement lifetime's dependence on $B$ found in \cite{briegel}, there is nothing limiting the slope $|dT_{E}/dB|$, i.e. by zooming into the region of the steep B-dependence, the slope increases arbitrarily, and hence the violation of \eqref{msens} is arbitrarily high.
We will now clarify the root of this inconsistency. Briegel and coworkers considered $T_{E}$ to be determined {\it just} by the magnetic interactions present in the radical-ion-pair. This is an erroneous association that disregards S-T decoherence inherent in radical-ion pairs. As in all precision measurements, the spin coherence time is the upper limit of the entanglement lifetime \cite{huelga,geremia,andre,lukin,budker}. Parenthetically, this is the reason why entangled states are hardly helpful in surpassing the standard quantum limits in quantum parameter estimation \cite{auzinsh,boixo}, i.e. any entanglement decays at least as fast as the coherent superpositions. In contrast, the authors in \cite{briegel} calculate an entanglement lifetime that bears no relation to the relevant decoherence time, which is on the order of $T_{r}$, in fact they find that in cases $T_{E}\gg T_{r}$.
\begin{figure}[h]
\includegraphics[width=8.5 cm]{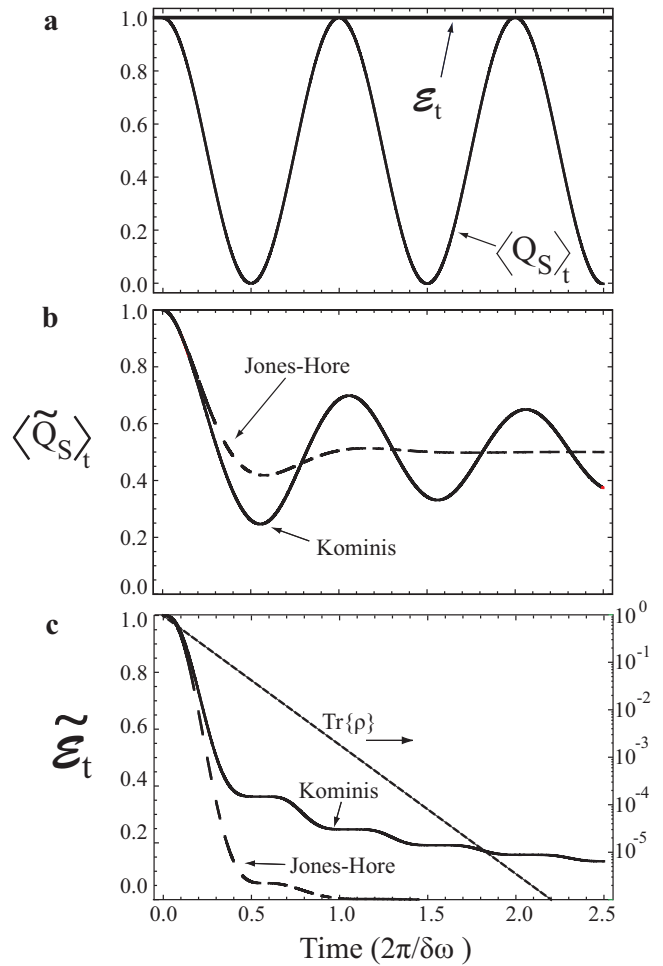}
\caption{(Color Online) (a) Time evolution of $\langle Q_{S}\rangle_{t}$ and ${\cal E}_{t}$ for the case of no reaction, $k_{S}=k_{T}=0$, calculated from $d\rho/dt=-i[{\cal H},\rho]$. (b) Time evolution of $\langle \tilde{Q}_{S}\rangle_{t}$  and (c) ${\cal \tilde{E}}_{t}$ for the realistic case of non-zero recombination rates $k_{S}=k_{T}=\delta\omega$, calculated with the full master equation $d\rho/dt=-i[{\cal H},\rho]+{\cal L}(\rho)$ of the Jones-Hore theory \cite{JH} and the Kominis theory \cite{kom2}. The full master equation of the traditional theory \cite{ivanov,purtov} leads to the same result as in case (a). In the right y-axis of (c) we plot the normalization of $\rho$, in order to elucidate the fact that there still exists a substantial number of radial-ion pairs at the time when the predictions of the two theories (Jones-Hore and Kominis) start to significantly deviate from the third (traditional). }
 \label{fig2}
\end{figure}
To further elucidate the root of the problem, we will consider a very simple example of a radical-ion pair without nuclear spins, just the two electrons, starting out in the singlet state $|\psi_{0}\rangle=|S\rangle=(|+-\rangle-|-+\rangle)/\sqrt{2}$. We suppose that the only magnetic interactions are the Zeeman interactions of the two electrons, and in order to induce S-T mixing, we consider the $\Delta g$ mechanism \cite{deltag}, i.e. the Larmor frequencies of the two electrons are taken to be slightly different. Thus the magnetic Hamiltonian is simply ${\cal H}=\omega_{1}s_{1z}+\omega_{2}s_{2z}$. If we completely disregard the reaction dynamics and for the moment assume that the evolution of the density matrix is driven solely by ${\cal H}$, we find the the initial state $|\psi_{0}\rangle$ evolves to $|\psi_{t}\rangle=e^{-i\delta\omega t/2}[(1+e^{i\delta\omega t})|S\rangle+(1-e^{i\delta\omega t})|T_{0}\rangle]/2$, where $|T_{0}\rangle=(|+-\rangle+|-+\rangle)/\sqrt{2}$ is the zero-projection state of the triplet manifold spanned by $|T_{0}\rangle$ and $|T_{\pm}\rangle=|\pm\pm\rangle$. As expected, ${\cal H}$ induces a coherent oscillation ${\rm S}\rightleftarrows{\rm T}_{0}$ at a frequency $\delta\omega=\omega_{1}-\omega_{2}$. It is easily shown that the concurrence \cite{wooters}, i.e. the overlap of $|\psi_{t}\rangle$ with the time-inverted $|\tilde{\psi}_{t}\rangle=\sigma_{y}|\psi^{*}_{t}\rangle$ is at all times $C_{t}=|\langle\psi_{t}|\tilde{\psi}_{t}\rangle|=1$. Thus the electron spin state starts out and remains maximally entangled and maximally coherent. In reality, however, radical-ion pairs suffer a continuous loss of S-T coherence due to the continuous quantum measurement induced by the recombination dynamics. In a single-molecule picture, this intramolecule measurement results in random quantum jumps either to the singlet or the triplet manifold of the RP spin states. In an ensemble of RPs this is equivalent to an improper mixture of singlet and triplet states, with the concomitant loss of S-T coherence {\it and} entanglement. The above considerations are illustrated in Fig. \ref{fig2}. 

In Fig. \ref{fig2}a we plot the expectation value of $Q_{S}$ and the entanglement ${\cal E}_{t}={\cal E}(C_{t}(\rho))$ \cite{wooters} for a fictitious radical-ion pair, for which $k_{S}=k_{T}=0$, i.e. when the evolution of $\rho$ is driven just by the magnetic Hamiltonian ${\cal H}$ and the singlet and triplet recombination channels simply do not exist. As explained above, the expectation value $\langle Q_{S}\rangle_{t}$ exhibits an undamped S-T oscillation with frequency $\delta\omega$, while the entanglement remains at its maximum value of one indefinitely, i.e. both the coherence and the entanglement lifetimes seem to be infinite. 
In Figs. \ref{fig2}b and \ref{fig2}c we consider the physical case, i.e. we turn on the recombination channels and take $k_{S}=k_{T}=\delta\omega$. Before explaining the result, we note that there currently exist three different theories attempting to describe the fundamental dynamics of radical-ion-pair reactions, the traditional theory of spin chemistry derived in \cite{ivanov,purtov}, a new theory based on quantum measurement considerations \cite{kom1,kom2} and a theory based on a different interpretation of the quantum measurement introduced in \cite{JH}. All three theories are of the form $d\rho/dt=-i[{\cal H},\rho]+{\cal L}(\rho)$, where $\rho$ is the electron and nuclear spin density matrix and ${\cal L}(\rho)$ the  super-operator describing the effect of the reaction. The three theories differ in the particular form of ${\cal L}(\rho)$, while they obviously agree in the imaginary scenario of no reaction, since when $k_{S}=k_{T}=0$ it is ${\cal L}(\rho)=0$, and their common result was depicted in Fig. \ref{fig2}a. Turning on the reaction channels, however, vast differences show up. In Figs. \ref{fig2}b and \ref{fig2}c it is seen that both S-T coherence and entanglement, respectively, decay in {\it both} the Jones-Hore and the Kominis theory, (albeit on a different time scale for each theory), exactly because the intra-molecule measurement dynamics transform coherent superpositions into incoherent mixtures. In contrast, the traditional theory leads to exactly the same result as in Fig. \ref{fig2}a, the case of no reaction! The reason is that the traditional theory is founded on the erroneous physical assumption that nothing  happens to surviving (unrecombined) radical-ion pairs. Indeed, since in the considered case of $k_{S},k_{T}\neq 0$ all radical-ion pairs eventually recombine away, i.e. $\tr\{\rho\}\rightarrow 0$ as $t\rightarrow\infty$, we normalize $\rho$ by $\tr\{\rho\}$, in order to describe the entanglement and singlet expectation value, as a function of time $t$, of those molecules that have not recombined until time $t$.  What is seen in Fig. \ref{fig2}b is that the oscillations of $\langle \tilde{Q}_{S}\rangle_{t}$ are damped away while $\langle \tilde{Q}_{S}\rangle_{t}$ decays towards $\langle \tilde{Q}_{S}\rangle_{\infty}$, where $\langle \tilde{Q}_{S}\rangle_{t}=\tr\{Q_{S}\rho\}/\tr\{\rho\}$. Similarly (Fig. \ref{fig2}c), $\tilde{\cal E}_{t}={\cal E}(C_{t}(\rho/\tr\{\rho\}))$ also decays to zero. Thus in reality, both the coherence and the entanglement lifetimes are finite and on the order of the reaction time.
\newline\indent
Put differently, the premise of the entanglement lifetime calculations of \cite{briegel}, a premise embedded in the traditional theory, is that {\it until they recombine}, RPs evolve unitarily under the action of ${\cal H}$ {\it only}. In contrast, the other two theories explicitly address  the fact that the recombination dynamics are not only responsible for transforming RPs into neutral products, but also for inducing S-T decoherence of RPs {\it until} they recombine. The absence of this physical mechanism from the fundamental theoretical description of RP reactions leads to {\it unphysical coherence and entanglement dynamics}, and in their turn, these lead to violation of fundamental quantum limits. Furthermore, the entanglement considerations of Vedral and coworkers \cite{vedral} are also based on the traditional theory, as the authors themselves state, rendering their qualitative and quantitative conclusions questionable.\newline\indent
In summary, we have set the entanglement dynamics of the chemical compass on a fundamentally sound basis. In so doing, we have shown that one out of the three currently competing master equations purporting to describe the quantum dynamics of radical-ion-pair reactions cannot stand as a fundamental theory, since it violates well-established fundamental quantum limits of precision measurements.

\end{document}